\documentclass[10pt,a4paper,oneside,notitlepage,fleqn,reqno,english,table]{article}

\usepackage[hidelinks]{hyperref} 

\usepackage{multicol} 

\usepackage{longtable} 
\usepackage{multirow} 
\usepackage{rotating} 
\usepackage[cp1252]{inputenc} 
\usepackage{lscape} 
\usepackage{url}    
\usepackage[section]{placeins}  
\usepackage{afterpage} 
\usepackage{capt-of}        
\usepackage{dirtytalk} 

\usepackage[T1]{fontenc}
\usepackage{tabularx, booktabs}   
\usepackage{makecell} 
\usepackage{array} 
\usepackage{xcolor} 
\usepackage{xfrac} 

\graphicspath{{figs/}} 					
\usepackage[parfill]{parskip} 					

\usepackage{amsmath} 			
\usepackage{amsfonts}			
\usepackage{mathtools}			
\everymath{\displaystyle}		
\usepackage{amssymb}  			
\usepackage{nicefrac}  			

\usepackage{tikz}  
\usetikzlibrary{shapes,arrows,calc,shadows,fit,intersections,datavisualization.formats.functions}
\tikzstyle{block}        = [draw, fill=blue!30, rectangle, text centered, minimum height=3em, minimum width=6em] 
\tikzstyle{title}        = [text centered]
\pgfdeclarelayer{background}
\pgfdeclarelayer{foreground}
\pgfsetlayers{background,main,foreground}

\usepackage{pgfplots}
\pgfplotsset{compat=1.13}
\pgfplotsset{colormap/bluered}
\usepackage{pgfplotstable}
\usepgfplotslibrary{polar}
\usepgfplotslibrary{colormaps}
\usepgfplotslibrary{external}

\usepackage{graphicx, color} 	

\allowdisplaybreaks 

\newcommand{\st}	{\scriptsize}           
\newcommand{\sss}   {\scriptscriptstyle}    

\newcommand{\nm}   		[1] {\ensuremath{\mathrm{#1}}} 									
\newcommand{\neweq}     [2] {\begin{equation} \mathrm{#1}\label{#2} \end{equation}} 	

\newcommand{\CC}		{{C\nolinebreak[4]\hspace{-.05em}\raisebox{.4ex}{\tiny\bf ++}}}

\renewcommand{\vec}		[1] {\mbox{\boldmath{\ensuremath{\mathrm{#1}}}}}	
\newcommand{\deriv}     [1] {\dfrac{d{#1}}{dt}}             				
\newcommand{\derpar}    [2] {\dfrac{d{#1}}{d{#2}}}      					
\newcommand{\pderpar}   [2] {\dfrac{\partial{#1}}{\partial{#2}}}    		
\newcommand{\lrp}       [1] {\left(#1\right)}								
\newcommand{\lrsb}      [1] {\left[#1\right]}								
\newcommand{\lrb}       [1] {\left\{#1\right\}}								

\newcommand{\FE}		 {F_{\sss E}}
\newcommand{\OECEF}    	 {O_{\sss E}}                           

\newcommand{\iEi}        {\vec i_{\sss1}^{\sss E}}
\newcommand{\iEii}       {\vec i_{\sss2}^{\sss E}}
\newcommand{\iEiii}      {\vec i_{\sss3}^{\sss E}}

\newcommand{\xEgdt}      {\vec x_{\sss GDT}}

\newcommand{\ON}          {O_{\sss N}}                            

\newcommand{\iNi}         {\vec i_{\sss1}^{\sss N}}
\newcommand{\iNii}        {\vec i_{\sss2}^{\sss N}}
\newcommand{\iNiii}       {\vec i_{\sss3}^{\sss N}}

\newcommand{\gc}     	  {\vec g_c}

\newcommand{\Hp}            {H_{\sss P}}

\newcommand{\DeltaT}        {\Delta T}
\newcommand{\Deltap}        {\Delta p}

\newcommand{\Tzero}         {T_{\sss 0}}                 
\newcommand{\pzero}         {p_{\sss 0}}                 
\newcommand{\rhozero}       {\rho_{\sss 0}}              
\newcommand{\THpzero}       {T_{\sss \Hp = 0}}           
\newcommand{\TISAHpzero}    {T_{\sss ISA,\Hp = 0}}       
\newcommand{\pHpzero}       {p_{\sss \Hp = 0}}           
\newcommand{\rhoHpzero}     {\rho_{\sss ISA,\Hp = 0}}	 
\newcommand{\HpHpzero}      {H_{\sss P,\Hp = 0}}     	 
\newcommand{\HHpzero}       {H_{\sss \Hp = 0}}           
\newcommand{\Hpzero}        {H_{\sss {P,\Hp = 0}}}       

\newcommand{\TMSL}          {T_{\sss MSL}}               
\newcommand{\TISAMSL}       {T_{\sss ISA,MSL}}           
\newcommand{\pMSL}          {p_{\sss MSL}}               
\newcommand{\HpMSL}         {H_{\sss P,MSL}}             
\newcommand{\HMSL}          {H_{\sss MSL}}               
\newcommand{\hMSL}          {h_{\sss MSL}}               

\newcommand{\TISA}          {T_{\sss ISA}}              
\newcommand{\TISAtrop}      {T_{\sss ISA,TROP}}         
\newcommand{\TISAbelow}     {T_{\sss ISA,<}} 
\newcommand{\TISAabove}     {T_{\sss ISA,>}} 

\newcommand{\betaT}         {\beta_{\sss T}}         	
\newcommand{\betaTbelow}    {\beta_{\sss T,<}}
\newcommand{\betaTabove}    {\beta_{\sss T,>}}

\newcommand{\Hptrop}        {H_{\sss P,TROP}}           
\newcommand{\Hpbelow}       {H_{\sss P,<}}
\newcommand{\Hpabove}       {H_{\sss P,>}}

\newcommand{\Ttrop}         {T_{\sss TROP}}             
\newcommand{\Tbelow}        {T_{\sss <}} 
\newcommand{\Tabove}        {T_{\sss >}} 

\newcommand{\ptrop}         {p_{\sss TROP}}             
\newcommand{\pbelow}        {p_{\sss <}} 
\newcommand{\pabove}        {p_{\sss >}} 

\newcommand{\Htrop}         {H_{\sss TROP}}             
\newcommand{\Hbelow}        {H_{\sss <}} 
\newcommand{\Habove}        {H_{\sss >}} 

\newcommand{\gBRbelow}      {\dfrac{- \gzero}{\betaTbelow \; R}}	
\newcommand{\gBRbelowinv}   {\dfrac{- \betaTbelow \; R}{\gzero}}	

\newcommand{\gzero}     	{g_{\sss 0}} 				
\newcommand{\RE}     	    {R_{\sss E}}   				

\newcommand{\pD}			{p_{\sss D}}
\newcommand{\TD}			{T_{\sss D}}
\newcommand{\TISAD}			{T_{\sss ISA,D}}
\newcommand{\lambdaD}		{\lambda_{\sss D}}
\newcommand{\phiD}			{\varphi_{\sss D}}
\newcommand{\tD}			{t_{\sss D}}
\newcommand{\hD}			{h_{\sss D}}
\newcommand{\HD}			{H_{\sss D}}
\newcommand{\HpD}			{H_{\sss P,D}}
\newcommand{\DeltaTD}       {\Delta T_{\sss D}}
\newcommand{\DeltapD}       {\Delta p_{\sss D}}
\newcommand{\TISAMSLD}      {T_{\sss ISA,MSL,D}}
\newcommand{\HpMSLD}        {H_{\sss P,MSL,D}}          
\newcommand{\pMSLD}         {p_{\sss MSL,D}}          

\begin{document}

\title{Quasi Static Atmospheric Model for Aircraft Trajectory Prediction and Flight Simulation}
\author{Eduardo Gallo\footnote{The author holds a MSc in Aerospace Engineering by the Polytechnic University of Madrid and has twenty-two years of experience working in aircraft trajectory prediction, modeling, and flight simulation. He is currently a Senior Trajectory Prediction and Aircraft Performance Engineer at Boeing Research \& Technology Europe (BR\&TE), although he is publishing this article in his individual capacity and time as part of his PhD thesis titled ``Autonomous Unmanned Air Vehicle GNSS-Denied Navigation'', advised by Dr. Antonio Barrientos within the Centre for Automation and Robotics of the Polytechnic University of Madrid.} \footnote{Contact: edugallo@yahoo.com, \url{https://orcid.org/0000-0002-7397-0425}}}
\date{January 2021}
\maketitle


\section*{Abstract}

Aircraft trajectory prediction requires the determination of the atmospheric properties (pressure, temperature, and density) encountered by the aircraft during its flight. This is accomplished by employing a tabulated prediction published by a meteorological service, a static atmosphere model that does not consider the atmosphere variation with time or horizontal position, such as the International Civil Aviation Organization (ICAO) Standard Atmosphere (ISA), or a variation to the later so it better conforms with the expected flight conditions. This article proposes an easy-to-use quasi static model that introduces temperature and pressure variations while respecting all the hypotheses of the ISA model, resulting in more realistic trajectory predictions as the obtained atmospheric properties bear a higher resemblance with the local conditions encountered during the flight. The proposed model relies on two parameters, the temperature and pressure offsets, and converges to the ISA model when both are zero. Expressions to obtain the atmospheric properties are established, and their dependencies with both parameters explained. The author calls this model INSA (ICAO Non Standard Atmosphere) and releases its \nm{\CC} implementation as open-source software \cite{Gallo2020_nonstandard_atmosphere}.


\section{Introduction - Influence of Atmosphere in Trajectory Prediction}\label{sec:Intro}

A six degrees of freedom flight simulation is composed of different modules, such as the aircraft performances, the navigation and control systems, the onboard sensors, the mission or flight plan, the flight deck inputs, and models for the wind and atmosphere. The accuracy of the resulting trajectory is based on the resemblance of the different modules to the real entities that they represent. In particular, the atmospheric conditions (pressure, temperature, and density) comprise a key factor in the resulting trajectory as they heavily influence the aircraft performances, both aerodynamic (lift, drag, moments of the control surfaces) and propulsive (power plant thrust and torque, fuel consumption). Different performances induce different climbing and descent angles, together with different optimum air speeds and altitudes, all of which result in different time and altitude profiles for the resulting trajectory.

A three degrees of freedom simulation is sufficient when it is not necessary to obtain a detailed description of the dynamics of the different flight maneuvers, as is generally the case for Air Traffic Management (ATM) applications. In this case the simulation modules are only the aircraft performances (lift, drag, thrust, and fuel consumption), the mission or flight plan, and models for the wind and atmosphere. 

A more efficient ATM system needs to be more predictable, allowing higher automation, which requires precise trajectory computation and hence an accurate aircraft performance model such as BADA \cite{Nuic2010}, together with an atmosphere model capable of closely representing the real atmospheric conditions encountered during the flight \cite{Gallo2006, Gallo2007}. The elevated uncertainty of the current ATM system is related to the aircraft intentions (mission or flight plan) and its performances, together with the wind and atmospheric predictions. ATM systems are evolving towards an operational concept known as Trajectory Based Operations (TBO) that aims to enable a more coordinated decision making process between the different stakeholders. Such systems will rely on human centered automation schemes with widespread use of Decision Support Tools (DSTs), each of which containing an accurate trajectory predictor \cite{Lopez2007, Ruiz2018}. An in-depth description of the components of trajectory prediction including the atmosphere model, together with their importance in the reduction of the ATM system uncertainty, is presented in \cite{Bronsvoort2013}.

One of the main challenges for the success of the emerging TBO concepts is the coordination between the onboard automation systems, such as the Flight Management System (FMS), and the different ground based DSTs; these systems each rely on their own trajectory predictors with its different components, some of which may be proprietary. The importance of synchronization between the trajectory predictions generated onboard the aircraft and on the ground for successful TBO operations is described in \cite{Bronsvoort2015}. Research suggests that the most important factor in the accuracy of the resulting trajectory is the meteorological data (wind and atmosphere), and in that regard ground based systems usually have access to more accurate predictions \cite{Swierstra2011}. However, nowadays the FMS is in command, which disconnects both the pilot and the ground controller from the direct control over the flight, even as ground based trajectory prediction is more accurate because it has access to more detailed and up to date wind and atmosphere predictions than those stored in the FMS \cite{Bronsvoort2014, Swierstra2019}. Communication and synchronization of wind and atmospheric predictions between the air and the ground is hence necessary for the successful interaction among the different DSTs of a TBO based ATM system.

This article focuses on how to effectively provide a flight simulation or trajectory predictor with atmospheric properties that, while capturing the intrinsic variations caused by altitude, also adapt to the temperature and pressure differences that exist between different Earth locations as well as those at the same location at different times. The proposed easy-to-use model also facilitates determining the influence of different atmospheric scenarios on the resulting trajectory, as it is possible to model the presence anywhere along the trajectory of warmer (colder) conditions, or that of a high (low) pressure front.

Meteorological services publish weather predictions \cite{NOAA,ECMWF} covering a given area for the next few days as tables providing the atmospheric properties based on time, longitude, latitude, and altitude; they also provide past data in the same format. This is adequate when simulating a given flight not long before its departure, or when replicating the exact conditions of a prior flight, but not too practical if the objective is to determine the influence of varying atmospheric conditions on the proposed mission.

The International Civil Aviation Organization (ICAO) Standard Atmosphere (ISA) \cite{ISA} is an atmosphere model created for the purposes of aircraft instrument calibration and aircraft performance rating. It is based on a series of hypotheses that capture the relationships among the different atmospheric variables, but it is by construction a static model intended for standardization and only captures the atmosphere variations with altitude, but not those with time and horizontal position. The ISA model is however widely employed for flight simulations and trajectory prediction. There also exist models based on similar hypotheses, such as the U.S. Standard Atmosphere \cite{USatm} or the Jet Standard Atmosphere, but they share the same shortcomings when applied to trajectory prediction.

This article describes an atmospheric model specifically designed for the requirements of trajectory prediction and flight simulation. It enables the introduction of user defined variations of temperature and pressure based on time and horizontal position, but respects all the hypotheses included in ISA, hence ensuring a realistic variation of the atmospheric properties with altitude. As it is based on the same hypotheses as ISA, the author has called it the ICAO Non Standard Atmosphere, or INSA.

After introducing a few basic concepts in section \ref{sec:Prelim}, section \ref{sec:Static_Quasi} provides a high level description of what comprises an atmospheric model and introduces the architecture of the proposed INSA model. The differences between the ISA and INSA models are described in section \ref{sec:Non_Standard}, which is followed by section \ref{sec:ISA} that lists the hypotheses on which ISA relies, which are shared by INSA. Section \ref{sec:Relationships} comprises the bulk of this article and contains the derivation of the INSA model equations. In section \ref{sec:Suggested} the author has added important suggestions on how to best employ the proposed model and adjust it to ground observations. Finally, the conclusions are provided in section \ref{sec:Conclusions}.


\section{Preliminary Concepts}\label{sec:Prelim}

Before focusing on the INSA atmosphere model itself, this section introduces several concepts that are required to properly understand the rest of the article:
\begin{itemize}

	\item The World Geodetic System 1984 (WGS84) \cite{WGS84} is the de facto standard for aircraft geopositioning and navigation, defining the WGS84 ellipsoid as the reference Earth surface.

	\item In a static atmosphere, the potential energy of an air particle is given by its gravity potential or sum of the potentials caused by the Earth gravitation plus that created by the Earth rotation around its axis \cite{ISA}. For aircraft positioning and navigation, the WGS84 ellipsoid can be considered as a geopotential surface\footnote{An isopotential or geopotential surface is that in which all its points have the same potential value.} and adopted as mean sea level (MSL).
\begin{figure}[h]
\centering
\begin{tikzpicture}[auto,>=latex',scale=1.2]
	\coordinate (origin) at (+0.0,+0.0);
	\coordinate (origin1) at (+0.0,+1.6);
	\coordinate (origin2) at (+0.0,+2.2);
	\coordinate (origin3) at (+0.0,-0.5);
			
	\filldraw [black] (origin) circle [radius=2pt] node [left=1pt] {\nm{\OECEF}};
	\draw [name path = ellhor] (origin) ellipse [x radius=4.0, y radius=1.65];
	\draw [name path = ellhor2, dashed] (origin1) ellipse [x radius=(4.0/1.19), y radius=(1.25/1.19)];
	\draw [name path = ellver] (origin) ellipse [x radius=1.3, y radius=3.0];
	\draw [name path = ellmain, very thick] (origin) ellipse [x radius=4.0, y radius=3.0];
	
	\draw [name path = iECEFi, name intersections={of=ellhor and ellver, by={int1,int2,int3,int4}}, very thick] [->] (origin) -- ($(origin) + 1.25*(int3)$) node [below=3pt] {\nm{\iEi}};
	\draw [name path = iECEFii, very thick] [->] (origin) -- (+4.5,+0.0) node [right=3pt]  {\nm{\iEii}};
	\draw [name path = iECEFiii, very thick] [->] (origin) -- (+0.0,+3.5) node [right=4pt]  {\nm{\iEiii}};
	\draw [dashed, ultra thin] (origin) -- (origin3);
	\filldraw [black] (origin3) circle [radius=1pt];
	
	\path [red, name path = path1] [-] (origin1) -- (+3.4,-0.6);
	\draw [dashed, ultra thin, name path = path12, name intersections={of=path1 and ellhor2, by={pointQ}}] [-] (origin3) -- (pointQ);
	\draw [dashed, ultra thin, name path = path11] [-] (origin1) -- (pointQ);
	\filldraw [black] (pointQ) circle [radius=1pt];
	
	\path [red, name path = path2] [-] (origin2) -- (+3.4,-0.0);
	\path [red, name path = path3] [-] (origin3) -- ($(origin3)!1.4!(pointQ)$);
	\path [name path = pathP, name intersections={of=path2 and path3, by={pointP}}] [->] (pointQ) -- (pointP);
	\path node at ($(pointP) + (0.2,0.3)$) {\nm{\ON}};
	
	\draw [dashed, ultra thin, name path = path21] [-] (origin2) -- (pointP);
	\filldraw [black] (pointP) circle [radius=1pt];
		
	\path [red, name path = path4] [-] (origin) -- (+3.4,-2.2);
	\path [red, name path = path5] [-] (pointQ) ++(0,-2) -- (pointQ);
	\draw [dashed, ultra thin, name intersections={of=path4 and path5, by={pointq}}] [-] (pointQ) -- (pointq);
	\path [red, name path = path6] [-] (pointP) ++(0,-2.5) -- (pointP);
	\draw [dashed, ultra thin, name path = path66, name intersections={of=path4 and path6, by={pointp}}] [-] (origin) -- (pointp);	
	\draw [dashed, ultra thin, name path = path22] [-] (pointp) -- (pointP);
	
	\path [red, name path = path7] [-] (-1.2,-0.7) -- (+1.0,-0.3);
	\draw [blue, name intersections={of=iECEFi and path7, by={lambda1}}, name intersections={of=path4 and path7, by={lambda2}}, very thick] [->] (lambda1) .. controls (-0.2,-0.65) and (0.15,-0.55) .. (lambda2) node [pos=0.7, below=0pt] {\nm{\lambda}};
			
	\path [red, name path = path8] [-] (+0.9,-0.8) -- (+0.85,+0.8);
	\draw [blue, name intersections={of=path66 and path8, by={phi1}}, name intersections={of=path12 and path8, by={phi2}}, very thick] [->] (phi1) .. controls (1.0,-0.4) and (1.0,-0.2) .. (phi2) node [pos=0.5, right=-2pt] {\nm{\varphi}};
	
	\draw [ultra thick] [->] (pointP) -- ($(pointP)!2.0!(pointQ)$) node [left =0pt]  {\nm{\iNiii}};
	\draw [ultra thick] [<-] (pointP) ++(0.9,0.2) node [above=0pt]  {\nm{\iNii}} -- (pointP);
	\draw [ultra thick] [<-] (pointP) ++(-0.4,+0.6) node [above=-2pt]  {\nm{\iNi}} -- (pointP);
		
	\draw [blue, ultra thick] [->] (pointQ) -- (pointP) node [pos=0.5, above left] {\nm{h}};
	
	\node [left of=origin, node distance=2.7cm] (Greenwich) {\st{Greenwich Mer.}};	
	\coordinate (origin3) at (+0.0,+1.2);
	\node [left of=origin3, node distance=5.2cm] (Ellipsoid) {\st{WGS84 Ell.}};	
	\coordinate (origin4) at (+0.0,-1.0);
	\node [left of=origin4, node distance=3.0cm] (Equator) {\st{Equator}};		
	
\end{tikzpicture}
\caption{The \texttt{NED} frame and geodetic coordinates}
\label{fig:RefSystems_N}
\end{figure}
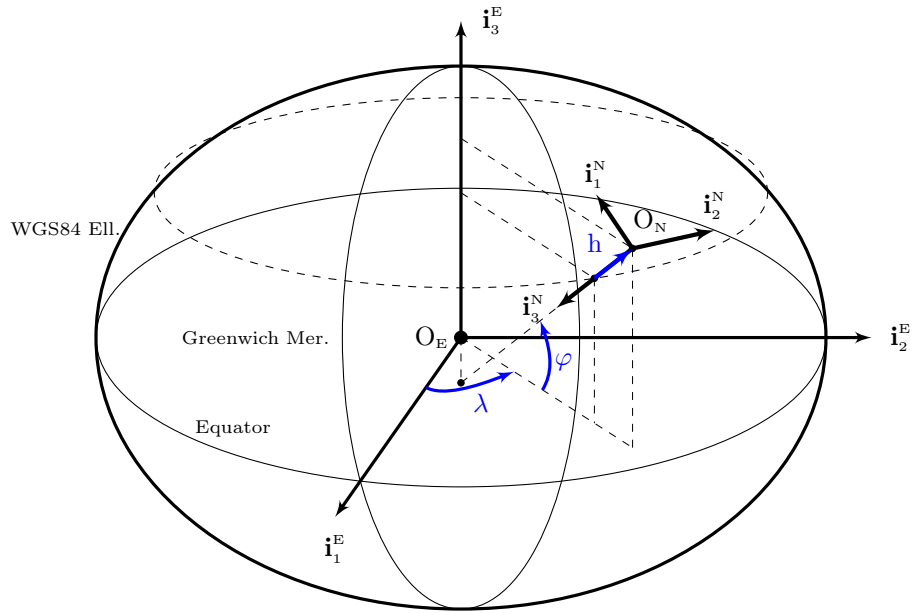

	\item Vectors representing different kinematic or dynamic aspects of the aircraft motion with respect to the Earth can be viewed in the Earth Centered Earth Fixed (ECEF) frame, a Cartesian reference system \nm{\FE = \lrb{\OECEF,\,\iEi,\,\iEii,\,\iEiii}} shown in figure \ref{fig:RefSystems_N}, where \nm{\OECEF} is located at the Earth center of mass (WGS84 ellipsoid center), \nm{\iEiii} points towards the geodetic North pole along the Earth rotation axis (ellipsoid symmetry axis), \nm{\iEi} is contained in both the Equator and Greenwich meridian planes pointing towards zero longitude, and \nm{\iEii} is orthogonal to \nm{\iEi} and \nm{\iEiii} forming a right hand system.

	\item The ECEF frame is also useful to define the geodetic coordinates \nm{\xEgdt = \lrsb{\lambda,\,\varphi,\,h}^T}, also shown in figure \ref{fig:RefSystems_N}, which enable the positioning of any point with respect to the Earth. \emph{Longitude} \nm{\lambda} \nm{\lrp{0 \leq \lambda < 2\pi }} is the angle formed between the Greenwich meridian plane (formed by \nm{\iEi} and \nm{\iEiii}) and the point meridian plane, \emph{latitude} \nm{\varphi} \nm{\lrp{-\pi/2 \leq \varphi \leq \pi/2 }} is the angle formed between the Equator plane and the line passing through the point that is orthogonal to the ellipsoid surface at the point where it intersects it, and \emph{geodetic altitude} h is the distance between the ellipsoid surface and the point measured along a line that is orthogonal to the ellipsoid surface (mean sea level MSL) at the point where it intersects it.
\neweq{\hMSL = 0}{eq:h_msl}

	\item \emph{Geopotential altitude} H is defined so that, when moving from a geopotential surface in a direction normal to its surface, the same differential work is performed by the gravity acceleration \nm{g_c} when displacing the unit of mass a geodetic distance dh as that performed by the standard acceleration of free fall \nm{\gzero} (table \ref{tab:constants}) when displacing the unit of mass a distance dH \cite{ISA}. The relationship between geopotential and geodetic altitudes is hence the following:
\neweq{- g_c \ dh = - g_0 \ dH}{eq:h_H_diff}

	Being a geopotential surface, mean sea level is not only taken as the reference for geodetic altitudes but also for geopotential ones \cite{ISA}:
\neweq{\HMSL = 0}{eq:H_msl}

	The relationship between the geodetic and geopotential altitudes is obtained by integrating (\ref{eq:h_H_diff}) between mean sea level (\nm{\hMSL = \HMSL = 0}) and a generic point.
\neweq{\int_{\hMSL=0}^{h} g_c \ dh = g_0 \int_{\HMSL=0}^{H} dH}{eq:h_H_intgr}

	As the gravity acceleration \nm{\gc} in an ellipsoidal Earth depends on both latitude and geodetic altitude, (\ref{eq:h_H_intgr}) lacks any explicit solution, although the results can be easily tabulated:
\begin{eqnarray}
	\nm{H} & = & \nm{f\lrp{h,\varphi}} \label{eq:h2H_table} \\
	\nm{h} & = & \nm{f\lrp{H,\varphi}} \label{eq:H2h_table}
\end{eqnarray}

	The influence of latitude is very small and the penalties for implementing a tabulated solution are significant, so it is common practice to employ a simplified solution obtained by solving (\ref{eq:h_H_intgr}) based on an spherical Earth surface of radius \nm{\RE} (table \ref{tab:constants}) \cite{ISA}, an spherical gravitation obtained by removing all terms except the first from the Earth Gravitational Model 1996 (\texttt{EGM96}) \cite{EGM96}, and an average centrifugal effect. This results in:
\begin{eqnarray}
	\nm{H} & = & \nm{\dfrac{\RE \cdot h}{\RE + h}} \label{eq:h2H} \\
	\nm{h} & = & \nm{\dfrac{\RE \cdot H}{\RE - H}} \label{eq:H2h}
\end{eqnarray}

\end{itemize}


\section{Static and Quasi Static Atmosphere Models} \label{sec:Static_Quasi}

When simulating the flight of an aircraft or predicting its trajectory, it is generally necessary to estimate the atmospheric properties (pressure p, temperature T, and density \nm{\rho}) of the air through which the aircraft flies, and sometimes also its derivatives with time. The atmospheric properties can vary with both time and geodetic position:
\neweq{\lrsb{p, \, T, \, \rho}^T = \vec f \lrp{t, \, \xEgdt} = \vec f \lrp{t, \, \lambda, \, \varphi, \, h}}{eq:model_generic_h}

Based on (\ref{eq:H2h}), the atmosphere model can also be expressed as:
\neweq{\lrsb{p, \, T, \, \rho}^T = \vec f \lrp{t, \, \lambda, \, \varphi, \, H}}{eq:model_generic_H}

An \emph{static atmosphere model} neglects the influence of time and horizontal position, and only provides the atmosphere dependency with geopotential altitude. ISA \cite{ISA} is such an static model.
\neweq{\lrsb{p, \, T, \, \rho}^T = \vec f \lrp{H}}{eq:model_H_static}

Static atmosphere models are too restrictive for flight simulation and trajectory prediction as they do not enable a proper representation of the variation of the atmosphere properties with time at a fixed location, or those between two different locations at the same time. A \emph{quasi static atmosphere model} acknowledges that the variations of atmospheric properties with time and horizontal position are much smaller than those with altitude,  and hence its time derivatives can be neglected:
\begin{eqnarray}
	\nm{\lrsb{p, \, T, \, \rho}^T} & = & \nm{\vec f \lrp{t, \, \lambda, \, \varphi, \, H}}\label{eq:model_H_quasi1} \\
	\nm{\deriv{\lrsb{p, \, T, \, \rho}^T}} & = & \nm{\deriv{\vec f \lrp{t, \, \lambda, \, \varphi, \, H}} \approx \pderpar{\vec f \lrp{t, \, \lambda, \, \varphi, \, H}}{H} \ \deriv{H}} \label{eq:model_H_quasi2}
\end {eqnarray}

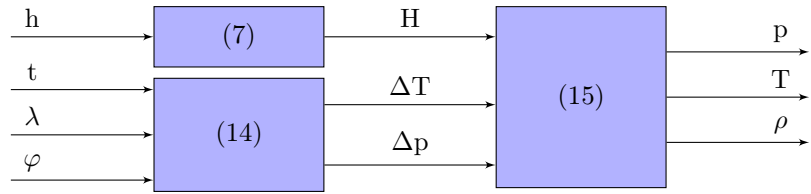
\begin{figure}[h]
\centering
\begin{tikzpicture}[auto, node distance=2cm,>=latex']
	\node [coordinate](lambdainput) {};
	\node [coordinate, above of=lambdainput, node distance=0.50cm] (refpoint){};
		
	\node [block, right of=lambdainput, text width=2.0cm, node distance=3.0cm, minimum height=1.5cm] (BLOCK1) {(\ref{eq:model_step1})};	
	\node [block, above of=BLOCK1, text width=2.0cm, node distance=1.3cm, minimum height=0.8cm] (BLOCK2) {(\ref{eq:h2H})};
	\node [block, right of=refpoint, text width=2.0cm, node distance=7.5cm, minimum height=2.4cm] (BLOCK3) {(\ref{eq:model_step2})};	

	\node [coordinate, above of=lambdainput, node distance=0.6cm] (tinput){};		
	\node [coordinate, below of=lambdainput, node distance=0.6cm] (phiinput){};
	\node [coordinate, above of=lambdainput, node distance=1.3cm] (hinput){};		
	\node [coordinate, right of=BLOCK3, node distance=3.0cm] (Toutput){};		
	\node [coordinate, above of=Toutput, node distance=0.6cm] (poutput){};	
	\node [coordinate, below of=Toutput, node distance=0.6cm] (rhooutput){};	

	\draw [->] (tinput) -- node[pos=0.15] {\nm{t}} ($(BLOCK1.west)+(0cm,0.6cm)$);
	\draw [->] (lambdainput) -- node[pos=0.15] {\nm{\lambda}} (BLOCK1.west);
	\draw [->] (phiinput) -- node[pos=0.15] {\nm{\varphi}} ($(BLOCK1.west)-(0cm,0.6cm)$);
	\draw [->] (hinput) -- node[pos=0.15] {\nm{h}} (BLOCK2.west);
	\draw [->] (BLOCK2.east) -- node[pos=0.5] {\nm{H}} ($(BLOCK3.west)+(0cm,0.8cm)$);
	\draw [->] ($(BLOCK1.east)+(0cm,0.4cm)$) -- node[pos=0.5] {\nm{\DeltaT}} ($(BLOCK3.west)-(0cm,0.1cm)$);
	\draw [->] ($(BLOCK1.east)-(0cm,0.4cm)$) -- node[pos=0.5] {\nm{\Deltap}} ($(BLOCK3.west)-(0cm,0.9cm)$);
	\draw [->] (BLOCK3.east) -- node[pos=0.8] {\nm{T}} (Toutput);
	\draw [->] ($(BLOCK3.east)+(0cm,0.6cm)$) -- node[pos=0.8] {\nm{p}} (poutput);
	\draw [->] ($(BLOCK3.east)-(0cm,0.6cm)$) -- node[pos=0.8] {\nm{\rho}} (rhooutput);
\end{tikzpicture}
\caption{Quasi static non standard atmosphere model flow diagram}
\label{fig:flow_diagram}
\end{figure}

This article proposes a quasi static atmosphere model called the ICAO Non Standard Atmosphere (INSA), represented in figure \ref{fig:flow_diagram}, that separates the computation of the atmospheric properties (\ref{eq:model_H_quasi1}) in two steps:
\begin{enumerate}
	\item The atmosphere variation with time and horizontal position is modeled through two parameters named the temperature and pressure offsets (\nm{\DeltaT}, \nm{\Deltap}):
	\neweq{\lrsb{\DeltaT, \, \Deltap}^T = \vec f_1 \lrp{t, \, \lambda, \, \varphi}}{eq:model_step1}
	\item At each time and horizontal position, the atmospheric properties are based on a static atmospheric model the complies with the ISA model \cite{ISA} hypotheses:
	\neweq{\lrsb{p, \, T, \, \rho}^T = \vec f_2 \lrp{H, \, \DeltaT, \, \Deltap}}{eq:model_step2}
\end{enumerate}

This model presents several advantages for flight simulation and trajectory prediction as it separates the weather (\ref{eq:model_step1}) from the local equilibrium of the atmosphere (\ref{eq:model_step2}). The former is not defined but allows the external user to employ any two three-dimensional functions \nm{ \vec f_1} to represent the weather conditions encountered throughout the specific trajectory being computed. The \emph{temperature offset} \nm{\DeltaT} represents the atmospheric temperature variations that occur daily\footnote{The temperature at a given location generally increases during the morning and descends during the evening.} or seasonally\footnote{The temperature at a given location of the Northern hemisphere is generally higher in summer and lower in winter.} at a single location, as well as those that occur at the same time at different latitudes\footnote{Locations near the poles are generally colder than those placed near the Equator.} or simply at different Earth locations, while the \emph{pressure offset} \nm{\Deltap} represents the atmospheric pressure changes due to the presence of high or low pressure weather systems. Note that none of these parameters model the diminution of atmospheric temperature and pressure with altitude, which is captured by (\ref{eq:model_step2}).

The static atmosphere model represented by (\ref{eq:model_step2}) provides the variation of the atmospheric properties with geopotential altitude along an infinitesimally narrow column of air normal to the Earth geopotential surfaces, which is uniquely defined by its temperature and pressure offsets. As the aircraft position changes during its flight, so do the offsets and hence the relationship between the atmospheric properties and geopotential altitude.


\section{Standard and Non Standard Static Atmosphere Models} \label{sec:Non_Standard}

The ICAO International Atmosphere model, also known as the \emph{standard atmosphere} or ISA, is a static model (\ref{eq:model_H_static}) defined by \cite{ISA} that provides expressions for the atmospheric pressure, standard temperature, and density as functions of the pressure altitude \nm{\Hp}:
\neweq{p, \, \TISA, \, \rho \, = f\,(\Hp)}{eq:model_ISA_standard}

Note that \emph{pressure altitude} \nm{\Hp} is defined as the geopotential altitude H that occurs in standard conditions, and \emph{standard temperature} \nm{\TISA} is the atmospheric temperature that would occur at a given pressure altitude if the conditions where those of the standard atmosphere. These two variables are also widely employed in non standard conditions below, but it is important to remark that in general \nm{\TISA \neq T} and \nm{\Hp \neq H}.

\emph{Standard conditions} are those provided by ISA (\ref{eq:model_ISA_standard}), while \emph{standard mean sea level conditions}, identified by the sub index ``\nm{\Hp = 0}'', correspond to their values where pressure altitude is zero (\nm{\Hpzero = 0}). They are \nm{\pHpzero = \pzero}, \nm{\TISAHpzero = \Tzero}, and \nm{\rhoHpzero = \rhozero}. The values of these constants are shown in table \ref{tab:constants}, which lists all the constants defined in \cite{ISA} that are employed throughout this article. Additionally, the ISA atmosphere is divided into two layers called the \emph{troposphere} below and the \emph{stratosphere} above; their separation is known as the \emph{tropopause}.
\begin{center}
\begin{tabular}{cp{6.5cm}rcc}
	\hline
	\textbf{Constant} & \textbf{Definition} 	& \textbf{Value} 		& \textbf{Units} \\
	\hline
	\nm{\gzero}      & standard acceleration of free fall     & 9.80665   & \nm{\lrsb{m / sec^{\sss2}}}       \\
	\nm{\RE}	     & Earth nominal radius                   & 6356766.0 & [m]                               \\
	\nm{\pzero}      & standard pressure at mean sea level    & 101325    & \nm{\lrsb{kg / m \, sec^{\sss2}}} \\
	\nm{\Tzero}      & standard temperature at mean sea level & 288.15    & \nm{\lrsb{^{\circ}K}}           \\
	\nm{\rhozero}    & standard density at mean sea level     & 1.225     & \nm{\lrsb{kg / m^{\sss3}}}      \\
	\nm{R}           & specific air constant                  & 287.05287 & \nm{\lrsb{m^{\sss2} / ^{\circ}K \; sec^{\sss2}}} \\
	\nm{\Hptrop}     & tropopause pressure altitude           & 11000     & [m] \\
	\nm{\betaTbelow} & temperature gradient below tropopause  & \nm{-6.5 \cdot 10^{\sss-3}} & \nm{\lrsb{^{\circ}K / m}} \\
	\nm{\betaTabove} & temperature gradient above tropopause  & 0         & \nm{\lrsb{^{\circ}K / m}} \\
	\hline
\end{tabular}
\end{center}
\captionof{table}{Constants defined by ISA \cite{ISA}} \label{tab:constants}

This article defines a \emph{non standard atmosphere} or INSA as a static model (\ref{eq:model_step2}) based on the same hypotheses as ISA but in which either \nm{\DeltaT} or \nm{\Deltap} is not zero (or both). Accordingly, \emph{non standard conditions} are those provided by an INSA model. The \emph{temperature offset} \nm{\DeltaT} and the \emph{pressure offset} \nm{\Deltap} are the differences in mean sea level conditions between a given INSA and ISA. \emph{Mean sea level} conditions, identified by the sub index ``MSL'', are those that occur where the geopotential altitude is zero (\nm{\HMSL = 0}), and differ from \nm{\lrp{\pHpzero, \, \TISAHpzero, \, \rhoHpzero}} in non standard conditions.

The limitations of using a static model such as ISA for trajectory prediction are obvious, and it is common to use a parameter similar to \nm{\DeltaT} to correct the temperature values provided by ISA, in what is known as ``OFF ISA'' conditions. Correcting for local pressure variations is far less common in ground based trajectory prediction, although the Query Nautical Height (QNH) and the Query Field Elevation (QFE) are employed to adjust the aircraft altimeters (and hence their FMS trajectory predictions) in the vicinity of airports\footnote{In this context, QNE refers to ISA conditions, and is only employed above a certain transition altitude.}. QNH represents the pressure altitude at mean sea level conditions (\nm{h = H = 0}), and QFE the pressure altitude at the airport.

All these ISA modifications are valid as long as the aircraft does not fly to far from the location where they apply because their values (\nm{\DeltaT}, QNH, and QFE) are constant; as such, they can only be employed for short trajectory segments (such as descents). The quasi static INSA model introduced in this article generalizes these modifications to the static ISA model into a comprehensive and easy-to-use scheme that enables continuous variations with time and horizontal position of the temperature and pressure differences with respect to ISA, while respecting all ISA hypotheses at the local level.
\begin{center}
\begin{tabular}{cclclclclclcl}
\hline
Variable & & \multicolumn{3}{c}{Standard Mean Sea Level} & & \multicolumn{3}{c}{Mean Sea Level} & & \multicolumn{3}{c}{Tropopause}\\
\hline
\nm{\Hp}   & \nm{\ \ } &  \nm{\Hpzero}     & = & \nm{0}     & \nm{\ \ } & \nm{\HpMSL}   &   & (\ref{eq:HpMSL})      & \nm{\ \ } & \nm{\Hptrop}   & = & \nm{11000 \lrsb{m}} \\
\nm{H}     & & \nm{\HHpzero}    &   & (\ref{eq:HHpzero})    &           & \nm{\HMSL}    & = & \nm{0}                &           & \nm{\Htrop}    &   & (\ref{eq:Htrop}) \\
\nm{p}     & & \nm{\pHpzero}    & = & \nm{\pzero}           &           & \nm{\pMSL}    & = & \nm{\pzero + \Deltap} &           & \nm{\ptrop}    &   & (\ref{eq:p_Hp_trop}) \\
\nm{\TISA} & & \nm{\TISAHpzero} & = & \nm{\Tzero}           &           & \nm{\TISAMSL} &   & (\ref{eq:TISAMSL})    &           & \nm{\TISAtrop} &   & (\ref{eq:TISA_Hp_trop}) \\
\nm{T}     & & \nm{\THpzero}    & = & \nm{\Tzero + \DeltaT} &           & \nm{\TMSL}    &   & (\ref{eq:TMSL})       &           & \nm{\Ttrop}    &   & (\ref{eq:T_Hp_trop}) \\
\hline
\end{tabular}
\end{center}
\captionof{table}{Standard mean sea level, mean sea level, and tropopause atmospheric conditions} \label{tab:smsl_msl}

Table \ref{tab:smsl_msl} shows the values of the atmospheric variables at standard mean sea level conditions (\nm{\Hp = 0}), mean sea level conditions (MSL), and at the tropopause (TROP), identifying whether the value corresponds to a definition obtained from ISA \cite{ISA} or the previous paragraphs, or whether it is obtained from an expression derived in this article.

A non standard atmosphere INSA is uniquely identified by its \nm{\DeltaT} and \nm{\Deltap} values, resulting in (\ref{eq:model_step2}). If both are zero, the non standard atmosphere converges to ISA. An INSA hence provides expressions for the atmospheric pressure, temperature, and density as functions of the geopotential altitude H and its two offset parameters (\ref{eq:model_step2}).


\section{Hypotheses of the ICAO Standard Atmosphere - ISA} \label{sec:ISA}

The non standard atmospheric model INSA proposed in this article is based on the same hypotheses as the ISA model, which are the following \cite{ISA}:
\begin{itemize}
	\item The law of perfect gases reflects the relationship between the atmospheric properties, where R (table \ref{tab:constants}) is the specific air constant.
\neweq{p = \rho \, R \, T}{eq:hyp_ideal_gas}

	\item Every air column corresponding to a given time and horizontal location is static in relation to the Earth surface, so the fluid equilibrium of forces in the direction normal to the Earth geopotential surfaces is the following:
\neweq{dp = - \rho \, g_c \, dh = - \rho \, \gzero \, dH}{eq:hyp_equilibrium}

	\item The tropopause altitude \nm{\Hptrop} (table \ref{tab:constants}) is constant when expressed in terms of pressure altitude.

	\item Each atmosphere layer is characterized by a constant temperature gradient with pressure altitude \nm{\betaT} (table \ref{tab:constants}), as shown in figure \ref{fig:dTdHp_Hp}:
\begin{figure}[h]
\centering
\begin{tikzpicture}
\begin{axis}[
cycle list={{blue,no markers,thick}},
width=16.0cm,
height=5.0cm,
xmin=0, xmax=15, xtick={0,1,...,15},
xlabel={\nm{\Hp \lrsb{km}}},
xmajorgrids,
ymin=-7, ymax=1, ytick={-7,-6,...,1},
ylabel={\nm{dT / d\Hp \lrsb{{^{\circ}K} / km}}},
ymajorgrids,
axis lines=left,
axis line style={-stealth},
]
\pgfplotstableread{figs/dTdHp_Hp.txt}\mytable
\addplot table [header=false, x index=0,y index=1] {\mytable};
\end{axis}   
\end{tikzpicture}
\caption{\nm{dT / d\Hp} versus \nm{\Hp}}
\label{fig:dTdHp_Hp}
\end{figure}
\neweq{\derpar{T}{\Hp} = \; \left\{
    \begin{array}{lcl}
        \nm{\betaTbelow} & \longrightarrow & \nm{\Hp <= \Hptrop} \\
        \nm{\betaTabove} & \longrightarrow & \nm{\Hp \ > \ \Hptrop}
    \end{array}
\right.}{eq:hyp_temp_gradient}
\end{itemize}


\section{Relationships among Atmospheric Variables} \label{sec:Relationships}

As indicated by (\ref{eq:model_step2}), a non standard atmosphere INSA identified by its temperature and pressure offsets (\nm{\DeltaT} and \nm{\Deltap}) defines the atmospheric pressure, temperature, and density as functions of geopotential altitude and the two offsets. These dependencies are repeated here for clarity:
\neweq{\lrsb{p, \, T, \, \rho}^T = \vec f_2 \lrp{H, \, \DeltaT, \, \Deltap}}{eq:model_step2_bis}

\begin{figure}[h]
\centering
\begin{tikzpicture}[auto, node distance=2cm,>=latex']
	\node [coordinate](Deltapinput) {};
		
	\node [block, right of=Deltapinput, text width=2.0cm, node distance=3.0cm, minimum height=1.5cm] (BLOCK4) {(\ref{eq:H_Hp_generic})};	
	\node [block, right of=Deltapinput, text width=2.0cm, node distance=7.5cm, minimum height=0.8cm] (BLOCK5) {(\ref{eq:p_Hp_generic})};	
	\node [block, above of=BLOCK5, text width=2.0cm, node distance=1.0cm, minimum height=0.8cm] (BLOCK6) {(\ref{eq:TISA_Hp_generic})};
	\node [block, below of=BLOCK5, text width=2.0cm, node distance=1.0cm, minimum height=0.8cm] (BLOCK7) {(\ref{eq:T_Hp__DeltaT_generic})};

	\node [coordinate, above of=Deltapinput, node distance=0.6cm] (Hinput){};		
	\node [coordinate, below of=Deltapinput, node distance=0.6cm] (DeltaTinput){};
	\node [coordinate, right of=DeltaTinput, node distance=1.0cm] (nodeDeltaT){};		
	\filldraw [black] (nodeDeltaT) circle [radius=1pt];
	\node [coordinate, right of=Deltapinput, node distance=5.6cm] (nodeHp){};	
	\filldraw [black] (nodeHp) circle [radius=1pt];
	\node [coordinate, right of=Deltapinput, node distance=10.5cm] (poutput){};
	\node [coordinate, above of=poutput, node distance=1.0cm] (Tisaoutput){};
	\node [coordinate, below of=poutput, node distance=1.0cm] (Toutput){};

	\draw [->] (Hinput) -- node[pos=0.15] {\nm{H}} ($(BLOCK4.west)+(0cm,0.6cm)$);
	\draw [->] (Deltapinput) -- node[pos=0.15] {\nm{\Deltap}} (BLOCK4.west);
	\draw [->] (DeltaTinput) -- node[pos=0.15] {\nm{\DeltaT}} ($(BLOCK4.west)-(0cm,0.6cm)$);
	\draw [->] (BLOCK4.east) -- node[pos=0.3] {\nm{\Hp}} (BLOCK5.west);
	\draw [->] (nodeHp) |- (BLOCK6.west);
	\draw [->] (nodeHp) |- ($(BLOCK7.west)+(0cm,0.25cm)$);
	\draw [->] (nodeDeltaT) |- ($(BLOCK7.west)-(0cm,0.25cm)$);
	\draw [->] (BLOCK5.east) -- node[pos=0.7] {\nm{p}} (poutput);
	\draw [->] (BLOCK6.east) -- node[pos=0.7] {\nm{\TISA}} (Tisaoutput);
	\draw [->] (BLOCK7.east) -- node[pos=0.7] {\nm{T}} (Toutput);
\end{tikzpicture}
\caption{Non standard atmosphere model flow diagram (static part)}
\label{fig:static_flow_diagram}
\end{figure}
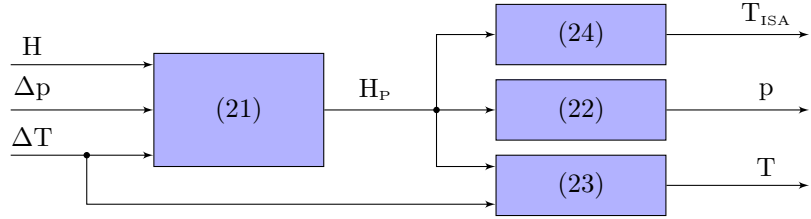

As shown in figure \ref{fig:static_flow_diagram}, the least complex way of obtaining these variables is to first obtain the pressure altitude \nm{\Hp} per (\ref{eq:H_Hp_generic}). Atmospheric pressure can then be obtained as a sole function of \nm{\Hp} (\ref{eq:p_Hp_generic}), while temperature also depends on \nm{\DeltaT} but not \nm{\Deltap} (\ref{eq:T_Hp__DeltaT_generic}). The standard temperature \nm{\TISA}, useful to simplify the previous expressions, is also a sole function of pressure altitude (\ref{eq:TISA_Hp_generic}). Density \nm{\rho} is finally obtained per (\ref{eq:hyp_ideal_gas}) as a function of pressure and temperature. The following sections derive these expressions from the hypotheses listed above.
\begin{eqnarray}
\nm{\Hp} & = & \nm{f \lrp{H, \, \DeltaT, \, \Deltap}}\label{eq:H_Hp_generic} \\
\nm{p} & = & \nm{f \lrp{\Hp}}\label{eq:p_Hp_generic} \\
\nm{T} & = & \nm{f \lrp{\Hp, \, \DeltaT}}\label{eq:T_Hp__DeltaT_generic} \\
\nm{\TISA} & = & \nm{f \lrp{\Hp}}\label{eq:TISA_Hp_generic}
\end{eqnarray}

\subsubsection*{Standard Temperature}

The standard temperature \nm{\TISA} is that occurring at a given pressure altitude \nm{\Hp} in standard conditions (\nm{\DeltaT = \Deltap = 0}). As indicated by (\ref{eq:TISA_Hp_generic}), it is a function of pressure altitude \nm{\Hp} exclusively. Its troposphere expression is obtained by integrating (\ref{eq:hyp_temp_gradient}) in standard conditions between standard mean sea level (\nm{\Hp = \Hpzero = 0}, \nm{\TISA = \TISAHpzero = \Tzero}) and any point below the tropopause.
\neweq{\TISAbelow = \Tzero + \betaTbelow \; \Hpbelow}{eq:TISA_Hp_below}
\begin{figure}[h]
\centering
\begin{tikzpicture}
\begin{axis}[
cycle list={{blue,no markers,thick}},
width=16.0cm,
height=5.0cm,
xmin=0, xmax=15, xtick={0,1,...,15},
xlabel={\nm{\Hp \lrsb{km}}},
xmajorgrids,
ymin=190, ymax=310, ytick={190,210,...,310},
ylabel={\nm{\TISA \lrsb{^{\circ}K}}},
ymajorgrids,
axis lines=left,
axis line style={-stealth},
]
\pgfplotstableread{figs/Tisa_Hp.txt}\mytable
\addplot table [header=false, x index=0,y index=1] {\mytable};
\end{axis}   
\end{tikzpicture}
\caption{\nm{\TISA} versus \nm{\Hp}}
\label{fig:Tisa_Hp}
\end{figure}

The stratosphere expression is obtained by integrating (\ref{eq:hyp_temp_gradient}) between the tropopause (\nm{\Hp = \Hptrop}, \nm{\TISA = \TISAtrop}) and any point above it, where \nm{\TISAtrop} is obtained from (\ref{eq:TISA_Hp_below}):
\begin{eqnarray}
\nm{\TISAtrop}  & = & \nm{\Tzero + \betaTbelow \; \Hptrop}\label{eq:TISA_Hp_trop} \\
\nm{\TISAabove} & = & \nm{\TISAtrop + \betaTabove \; \lrp{\Hpabove - \Hptrop} = \TISAtrop}\label{eq:TISA_Hp_above}
\end{eqnarray}

Note that (\ref{eq:TISA_Hp_generic}), which is a combination of (\ref{eq:TISA_Hp_below}) and (\ref{eq:TISA_Hp_above}), can not be reversed because of the constant standard temperature in the stratosphere. The relationship between \nm{\TISA} and \nm{\Hp}, graphically represented in figure \ref{fig:Tisa_Hp}, is valid for all INSA non standard atmospheres as it does not depend on \nm{\DeltaT} or \nm{\Deltap}.

The standard temperature at mean sea level \nm{\TISAMSL} shown in table \ref{tab:smsl_msl} is obtained by inserting the mean sea level pressure altitude \nm{\HpMSL} obtained in (\ref{eq:HpMSL}) into (\ref{eq:TISA_Hp_below}), resulting in:
\neweq{\TISAMSL = \Tzero + \betaTbelow \; \HpMSL}{eq:TISAMSL}


\subsubsection*{Temperature}

Temperature T is a function of both pressure altitude \nm{\Hp} and temperature offset \nm{\DeltaT} (\ref{eq:T_Hp__DeltaT_generic}). The expression for the troposphere is also obtained by integrating (\ref{eq:hyp_temp_gradient}), represented in figure \ref{fig:dTdHp_Hp}, between standard mean sea level (\nm{\Hp = \Hpzero = 0, \, T = \THpzero = \Tzero + \DeltaT}), where the temperature is taken from the temperature offset definition, and any point below the tropopause:
\neweq{\Tbelow = \Tzero + \DeltaT + \betaTbelow \; \Hpbelow = \TISAbelow + \DeltaT}{eq:T_Hp_below}

The stratosphere expression is obtained by integrating (\ref{eq:hyp_temp_gradient}) between the tropopause (\nm{\Hp = \Hptrop}, \nm{T = \Ttrop}) and any point above it, where \nm{\Ttrop} is obtained from (\ref{eq:T_Hp_below}):
\begin{eqnarray}
\nm{\Ttrop}  & = & \nm{\Tzero + \DeltaT + \betaTbelow \; \Hptrop = \TISAtrop + \DeltaT}\label{eq:T_Hp_trop} \\
\nm{\Tabove} & = & \nm{\Ttrop + \betaTabove \; \lrp{\Hpabove - \Hptrop} = \Ttrop}\label{eq:T_Hp_above}
\end{eqnarray}

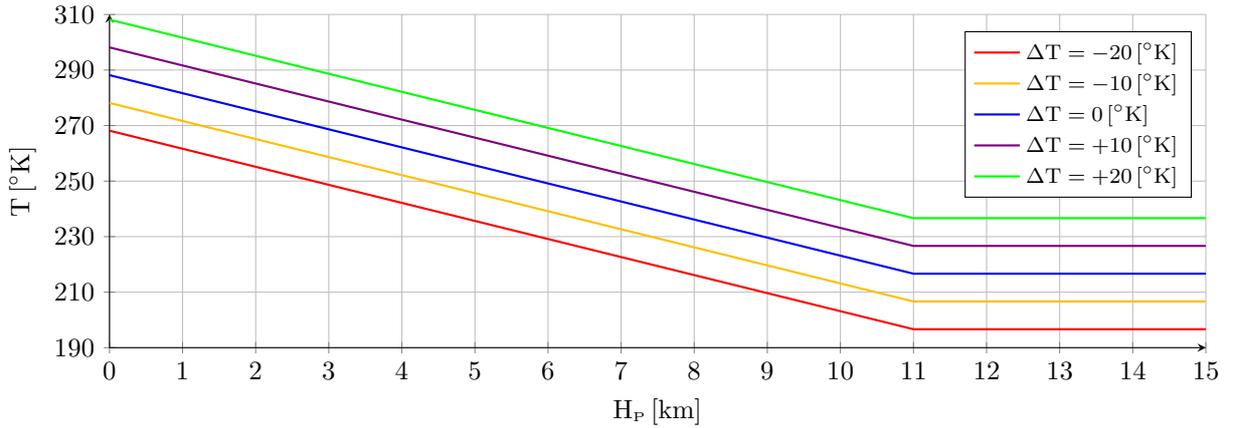
\begin{figure}[h]
\centering
\pgfplotsset{
	every axis legend/.append style={
		at={(0.78,0.70)},
		anchor=west,
	},
}
\begin{tikzpicture}
\begin{axis}[
cycle list={{red,no markers,thick},{orange!50!yellow,no markers,thick},{blue,no markers,thick},{violet,no markers,thick},{green,no markers,thick}},
width=16.0cm,
height=6.0cm,
xmin=0, xmax=15, xtick={0,1,...,15},
xlabel={\nm{\Hp \lrsb{km}}},
xmajorgrids,
ymin=190, ymax=310, ytick={190,210,...,310},
ylabel={\nm{T \lrsb{^{\circ}K}}},
ymajorgrids,
axis lines=left,
axis line style={-stealth},
legend entries={
\nm{\DeltaT = - 20 \lrsb{^{\circ}K}}, 
\nm{\DeltaT = - 10 \lrsb{^{\circ}K}}, 
\nm{\DeltaT = 0 \lrsb{^{\circ}K}}, 
\nm{\DeltaT = + 10 \lrsb{^{\circ}K}}, 
\nm{\DeltaT = + 20 \lrsb{^{\circ}K}}},
legend columns=1,
legend style={font=\footnotesize},
legend cell align=left,
]
\pgfplotstableread{figs/T_Hp__DeltaT.txt}\mytable
\addplot table [header=false, x index=0,y index=1] {\mytable};
\addplot table [header=false, x index=0,y index=2] {\mytable};
\addplot table [header=false, x index=0,y index=3] {\mytable};
\addplot table [header=false, x index=0,y index=4] {\mytable};
\addplot table [header=false, x index=0,y index=5] {\mytable};
\end{axis}   
\end{tikzpicture}
\caption{T versus \nm{\Hp} for various \nm{\DeltaT}}
\label{fig:T_Hp__DeltaT}
\end{figure}

Note that (\ref{eq:T_Hp__DeltaT_generic}), which is a combination of (\ref{eq:T_Hp_below}) and (\ref{eq:T_Hp_above}), can not be reversed because of the constant temperature in the stratosphere. The relationship between \nm{T} and \nm{\Hp} is graphically represented in figure \ref{fig:T_Hp__DeltaT} for various temperature offsets \nm{\DeltaT} as it does not depend on the pressure offset \nm{\Deltap}.

The temperature at mean sea level \nm{\TMSL} shown in table \ref{tab:smsl_msl} is obtained by inserting the mean sea level pressure altitude \nm{\HpMSL} obtained in (\ref{eq:HpMSL}) into (\ref{eq:T_Hp_below}), resulting in:
\neweq{\TMSL = \Tzero + \DeltaT +  \betaTbelow \; \HpMSL}{eq:TMSL}


\subsubsection*{Pressure}

The relationship between pressure p and pressure altitude \nm{\Hp} does not depend on either the temperature or pressure offsets (\ref{eq:p_Hp_generic}). The ratio between differential increments of pressure and pressure altitude is obtained by combining the law of perfect gases (\ref{eq:hyp_ideal_gas}) with the atmosphere fluid equilibrium (\ref{eq:hyp_equilibrium}) in standard conditions (\nm{\Hp = H} and \nm{\TISA = T}), resulting in:
\neweq{\frac{dp}{p} = -\frac{\gzero}{R \ \TISA} \ d\Hp}{eq:dp_dHp}
\begin{figure}[h]
\centering
\begin{tikzpicture}
\begin{axis}[
cycle list={{blue,no markers,thick}},
width=16.0cm,
height=5.0cm,
xmin=0, xmax=15, xtick={0,1,...,15},
xlabel={\nm{\Hp \lrsb{km}}},
xmajorgrids,
ymin=-12, ymax=0, ytick={-12,-10,...,0},
ylabel={\nm{dp / d\Hp \lrsb{kg / m^{\sss2} \, sec^{\sss2}}}},
ymajorgrids,
axis lines=left,
axis line style={-stealth},
]
\pgfplotstableread{figs/dpdHp_Hp.txt}\mytable
\addplot table [header=false, x index=0,y index=1] {\mytable};
\end{axis}   
\end{tikzpicture}
\caption{\nm{dp / d\Hp} versus \nm{\Hp}}
\label{fig:dpdHp_Hp}
\end{figure}

Below the tropopause, expression (\ref{eq:TISA_Hp_below}) can be inserted into (\ref{eq:dp_dHp}), resulting in:
\neweq{\frac{d\pbelow}{\pbelow} = -\frac{\gzero}{R} \ \frac{d\Hpbelow}{\Tzero + \betaTbelow \; \Hpbelow}}{eq:dp_dHp_below}

Its integration between standard mean sea level conditions (\nm{\Hp = \Hpzero = 0}, \nm{p = \pHpzero = \pzero}) and any point below the tropopause results in:
\neweq{\pbelow = \pzero \ \lrp{1 + \frac{\betaTbelow}{\Tzero} \ \Hpbelow}^{\sss \gBRbelow}}{eq:p_Hp_below}

In the case of the stratosphere, inserting (\ref{eq:TISA_Hp_above}) into (\ref{eq:dp_dHp}) results in:
\neweq{\frac{d\pabove}{\pabove} = -\frac{\gzero}{R} \ \frac{d\Hpabove}{\TISAtrop}}{eq:dp_dHp_above}

This expression can then be integrated between the tropopause (\nm{\Hp = \Hptrop}, \nm{p = \ptrop}) and any point above it, where \nm{\ptrop} is obtained from (\ref{eq:p_Hp_below}):
\begin{eqnarray}
\nm{\ptrop} & = & \nm{\pzero \ \lrp{1 + \frac{\betaTbelow}{\Tzero} \ \Hptrop}^{\sss \gBRbelow}}\label{eq:p_Hp_trop} \\
\nm{\pabove} & = & \nm{\ptrop \ exp \lrsb{\frac{- \gzero}{R \; \TISAtrop} \lrp{\Hpabove - \Hptrop}}}\label{eq:p_Hp_above}
\end{eqnarray}
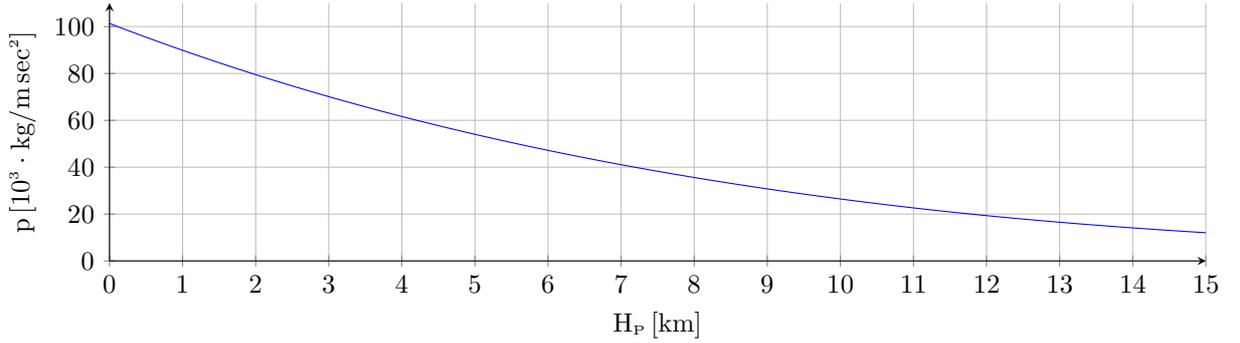
\begin{figure}[h]
\centering
\begin{tikzpicture}
\begin{axis}[
cycle list={{blue,no markers,thick}},
width=16.0cm,
height=5.0cm,
xmin=0, xmax=15, xtick={0,1,...,15},
xlabel={\nm{\Hp \lrsb{km}}},
xmajorgrids,
ymin=0, ymax=110, ytick={0,20,...,100},
ylabel={\nm{p \lrsb{10^{\sss3} \cdot kg / m \, sec^{\sss2}}}},
ymajorgrids,
axis lines=left,
axis line style={-stealth},
]
\pgfplotstableread{figs/p_Hp.txt}\mytable
\addplot table [header=false, x index=0,y index=1] {\mytable};
\end{axis}   
\end{tikzpicture}
\caption{p versus \nm{\Hp}}
\label{fig:p_Hp}
\end{figure}

The relationship between \nm{p} and \nm{\Hp}, graphically represented in figure \ref{fig:p_Hp}, is valid for all INSA non standard atmospheres as it does not depend on \nm{\DeltaT} or \nm{\Deltap}. Note that the decrease in atmospheric pressure with pressure altitude is slower the higher the pressure altitude.

Expressions (\ref{eq:p_Hp_below}) and (\ref{eq:p_Hp_above}) are easily reversed, resulting in:
\begin{eqnarray}
\nm{\Hpbelow} & = & \nm{\frac{\Tzero}{\betaTbelow} \; \lrsb{\lrp{\frac{\pbelow}{\pzero}}^{\sss \gBRbelowinv} - 1}}\label{eq:Hp_p_below} \\
\nm{\Hpabove} & = & \nm{\Hptrop - \frac{R \; \TISAtrop}{\gzero} \; \log_n{\lrp{\frac{\pabove}{\ptrop}}}}\label{eq:Hp_p_above}
\end{eqnarray}

The mean sea level pressure altitude \nm{\HpMSL} is obtained by replacing \nm{\pbelow} by \nm{\pMSL = \pzero + \Deltap} within (\ref{eq:Hp_p_below}), where \nm{\pMSL} is taken from its table \ref{tab:smsl_msl} definition:
\neweq{\HpMSL = \frac{\Tzero}{\betaTbelow} \; \lrsb{\lrp{\frac{\pMSL}{\pzero}}^{\sss \gBRbelowinv} - 1}}{eq:HpMSL}


\subsubsection*{Geopotential Altitude}

The combination of the law of perfect gases (\ref{eq:hyp_ideal_gas}) and the fluid static atmosphere vertical equilibrium (\ref{eq:hyp_equilibrium}) results in:
\neweq{\frac{dp}{p} = -\frac{\gzero}{R \ T} \ dH}{eq:dp_dH}

\begin{figure}[h]
\centering
\pgfplotsset{
	every axis legend/.append style={
		at={(0.02,1.11)},
		anchor=west,
	},
}
\begin{tikzpicture}
\begin{axis}[
cycle list={{red,no markers,thick},{orange!50!yellow,no markers,thick},{blue,no markers,thick},{violet,no markers,thick},{green,no markers,thick}},
width=16.0cm,
height=6.0cm,
xmin=0, xmax=15, xtick={0,1,...,15},
xlabel={\nm{\Hp \lrsb{km}}},
xmajorgrids,
ymin=0.90, ymax=1.10, ytick={0.90,0.95,1,1.05,1.10},
ylabel={\nm{dH / d\Hp \lrsb{-}}},
ymajorgrids,
axis lines=left,
axis line style={-stealth},
legend entries={
\nm{\DeltaT = - 20 \lrsb{^{\circ}K}}, 
\nm{\DeltaT = - 10 \lrsb{^{\circ}K}}, 
\nm{\DeltaT = 0 \lrsb{^{\circ}K}}, 
\nm{\DeltaT = + 10 \lrsb{^{\circ}K}}, 
\nm{\DeltaT = + 20 \lrsb{^{\circ}K}}},
legend columns=5,
legend style={font=\footnotesize},
legend cell align=left,
]
\pgfplotstableread{figs/dHdHp_Hp__DeltaT.txt}\mytable
\addplot table [header=false, x index=0,y index=1] {\mytable};
\addplot table [header=false, x index=0,y index=2] {\mytable};
\addplot table [header=false, x index=0,y index=3] {\mytable};
\addplot table [header=false, x index=0,y index=4] {\mytable};
\addplot table [header=false, x index=0,y index=5] {\mytable};
\end{axis}   
\end{tikzpicture}
\caption{\nm{dH / d\Hp} versus \nm{\Hp} for various \nm{\DeltaT}}
\label{fig:dHdHp_Hp__DeltaT}
\end{figure}
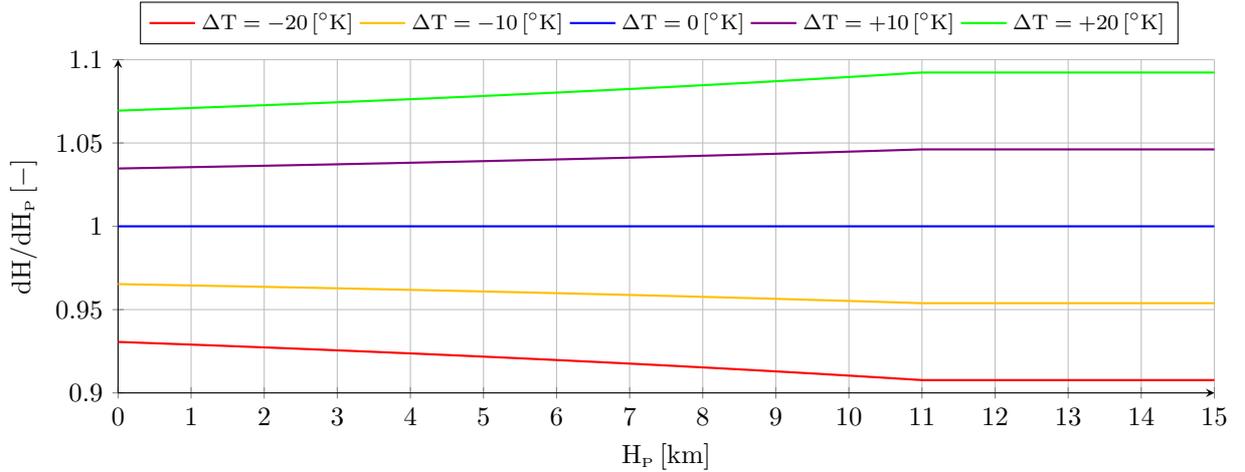

Dividing (\ref{eq:dp_dH}) by (\ref{eq:dp_dHp}) results in the ratio between differential changes in geopotential and pressure altitudes, which is equal to that between the atmospheric temperature and the temperature that would occur at the same pressure altitude if the atmospheric conditions were standard. This ratio is represented in figure \ref{fig:dHdHp_Hp__DeltaT}. Note that geopotential altitude H grows quicker than pressure altitude \nm{\Hp} in warm atmospheres (\nm{\DeltaT > 0}), and slower in cold ones.
\neweq{\frac{dH}{d\Hp} = \frac{T}{\TISA}}{eq:dH_dHp}

In the troposphere, the introduction of (\ref{eq:TISA_Hp_below}) and (\ref{eq:T_Hp_below}) into (\ref{eq:dH_dHp}) results in:
\neweq{\derpar{\Hbelow}{\Hpbelow} = \frac{\Tzero + \DeltaT + \betaTbelow \; \Hpbelow}{\Tzero + \betaTbelow \; \Hpbelow} = 1 + \frac{\DeltaT}{\Tzero + \betaTbelow \; \Hpbelow}}{eq:dh_dHp_below}

Its integration between mean sea level conditions (\nm{\Hp = \HpMSL}, \nm{H = \HMSL = 0}) and any point below the tropopause results in:
\neweq{\Hbelow = \Hpbelow - \HpMSL + \frac{\DeltaT}{\betaTbelow} \; \log_n \lrp{\frac{\Tzero + \betaTbelow \; \Hpbelow}{\TISAMSL}}}{eq:H_Hp_below}

where \nm{\HpMSL} is given by (\ref{eq:HpMSL}) and \nm{\TISAMSL} by (\ref{eq:TISAMSL}). In the stratosphere, the introduction of (\ref{eq:TISA_Hp_above}) and (\ref{eq:T_Hp_above}) into (\ref{eq:dH_dHp}) results in:
\neweq{\derpar{\Habove}{\Hpabove} = \frac{\Tzero + \DeltaT + \betaTbelow \; \Hptrop}{\Tzero + \betaTbelow \; \Hptrop} = \frac{\Ttrop}{\TISAtrop} = 1 + \frac{\DeltaT}{\Tzero + \betaTbelow \; \Hptrop}}{eq:dh_dHp_above}

Expression (\ref{eq:dh_dHp_above}) can then be integrated between the tropopause (\nm{\Hp = \Hptrop}, \nm{H = \Htrop}) and any point above it, where \nm{\Htrop} is obtained from (\ref{eq:H_Hp_below}):
\begin{eqnarray}
\nm{\Htrop} & = & \nm{\Hptrop - \HpMSL + \frac{\DeltaT}{\betaTbelow} \; \log_n \lrp{\frac{\Tzero + \betaTbelow \; \Hptrop}{\TISAMSL}}}\label{eq:Htrop} \\
\nm{\Habove} & = & \nm{\Htrop + \frac{\Ttrop}{\TISAtrop} \; \lrp{\Hpabove - \Hptrop}} \label{eq:H_Hp_above}
\end{eqnarray}

The standard mean sea level geopotential altitude \nm{\HHpzero}, shown in table \ref{tab:smsl_msl}, is obtained by replacing \nm{\Hpbelow} with \nm{\HpHpzero = 0} in (\ref{eq:H_Hp_below}), resulting in:
\neweq{\HHpzero = - \HpMSL + \frac{\DeltaT}{\betaTbelow} \; \log_n \lrp{\frac{\Tzero}{\TISAMSL}}}{eq:HHpzero}
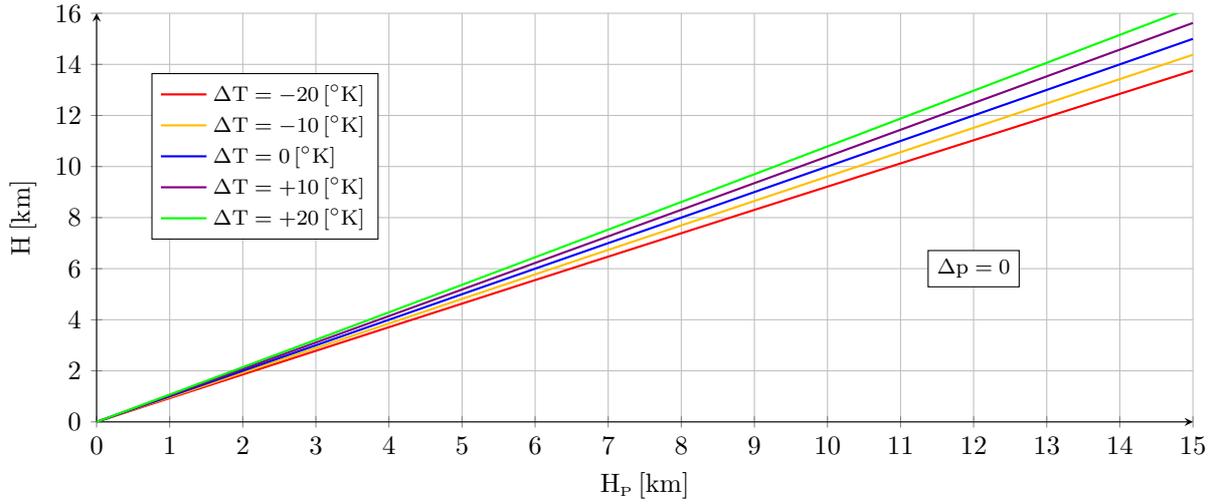
\begin{figure}[h]
\centering
\pgfplotsset{
	every axis legend/.append style={
		at={(0.05,0.65)},
		anchor=west,
	},
}
\begin{tikzpicture}
\begin{axis}[
cycle list={{red,no markers,thick},{orange!50!yellow,no markers,thick},{blue,no markers,thick},{violet,no markers,thick},{green,no markers,thick}},
width=16.0cm,
height=7.0cm,
xmin=0, xmax=15, xtick={0,1,...,15},
xlabel={\nm{\Hp \lrsb{km}}},
xmajorgrids,
ymin=0, ymax=16, ytick={0,2,...,16},
ylabel={\nm{H \lrsb{km}}},
ymajorgrids,
axis lines=left,
axis line style={-stealth},
legend entries={
\nm{\DeltaT = - 20 \lrsb{^{\circ}K}}, 
\nm{\DeltaT = - 10 \lrsb{^{\circ}K}}, 
\nm{\DeltaT = 0 \lrsb{^{\circ}K}}, 
\nm{\DeltaT = + 10 \lrsb{^{\circ}K}}, 
\nm{\DeltaT = + 20 \lrsb{^{\circ}K}}},
legend columns=1,
legend style={font=\footnotesize},
legend cell align=left,
]
\pgfplotstableread{figs/H_Hp__DeltaT.txt}\mytable
\addplot table [header=false, x index=0,y index=1] {\mytable};
\addplot table [header=false, x index=0,y index=2] {\mytable};
\addplot table [header=false, x index=0,y index=3] {\mytable};
\addplot table [header=false, x index=0,y index=4] {\mytable};
\addplot table [header=false, x index=0,y index=5] {\mytable};
\path node [draw, shape=rectangle, fill=white] at (12.0,6.0) {\footnotesize \nm{\Deltap = 0}};
\end{axis}   
\end{tikzpicture}
\caption{H versus \nm{\Hp} at \nm{\Deltap = 0} for various \nm{\DeltaT}}
\label{fig:H_Hp__DeltaT}
\end{figure}

Note that while both \nm{\DeltaT} and \nm{\Deltap} influence the relationship between H and \nm{\Hp} (\ref{eq:H_Hp_generic}), they do so in different ways. While \nm{\DeltaT} sets the ratio between increments of both types of altitudes, the influence of \nm{\Deltap} is by means of the mean sea level pressure \nm{\pMSL}. In other words, when representing H versus \nm{\Hp}, the temperature offset \nm{\DeltaT} sets the slope and the pressure offset \nm{\Deltap} marks the zero point.
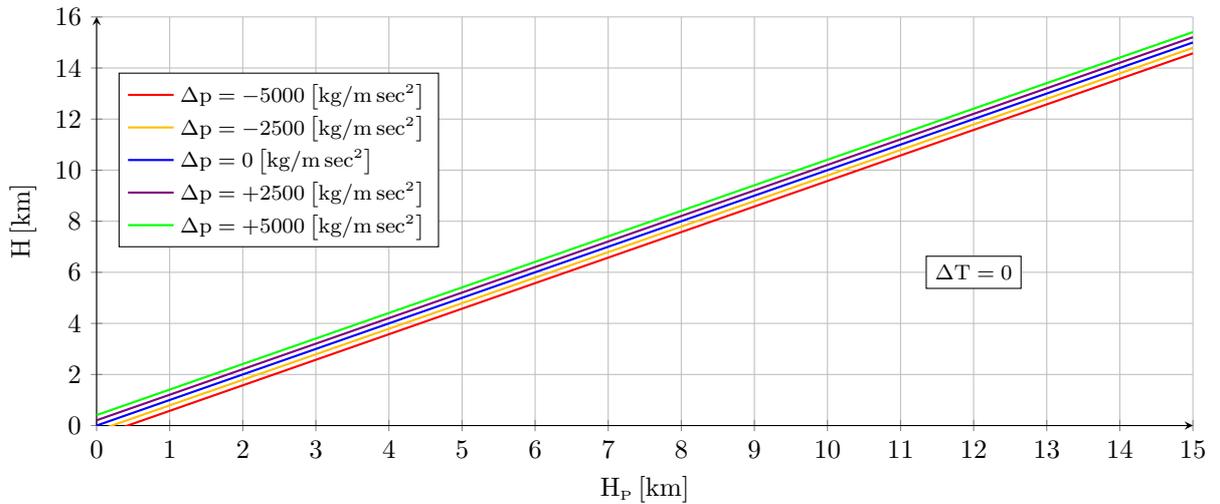
\begin{figure}[h]
\centering
\pgfplotsset{
	every axis legend/.append style={
		at={(0.02,0.65)},
		anchor=west,
	},
}
\begin{tikzpicture}
\begin{axis}[
cycle list={{red,no markers,thick},{orange!50!yellow,no markers,thick},{blue,no markers,thick},{violet,no markers,thick},{green,no markers,thick}},
width=16.0cm,
height=7.0cm,
xmin=0, xmax=15, xtick={0,1,...,15},
xlabel={\nm{\Hp \lrsb{km}}},
xmajorgrids,
ymin=0, ymax=16, ytick={0,2,...,16},
ylabel={\nm{H \lrsb{km}}},
ymajorgrids,
axis lines=left,
axis line style={-stealth},
legend entries={
\nm{\Deltap = - 5000 \lrsb{kg / m \, sec^{\sss2}}}, 
\nm{\Deltap = - 2500 \lrsb{kg / m \, sec^{\sss2}}},
\nm{\Deltap = 0 \lrsb{kg / m \, sec^{\sss2}}},
\nm{\Deltap = + 2500 \lrsb{kg / m \, sec^{\sss2}}},
\nm{\Deltap = + 5000 \lrsb{kg / m \, sec^{\sss2}}}},
legend columns=1,
legend style={font=\footnotesize},
legend cell align=left,
]
\pgfplotstableread{figs/H_Hp__Deltap.txt}\mytable
\addplot table [header=false, x index=0,y index=1] {\mytable};
\addplot table [header=false, x index=0,y index=2] {\mytable};
\addplot table [header=false, x index=0,y index=3] {\mytable};
\addplot table [header=false, x index=0,y index=4] {\mytable};
\addplot table [header=false, x index=0,y index=5] {\mytable};
\path node [draw, shape=rectangle, fill=white] at (12.0,6.0) {\footnotesize \nm{\DeltaT = 0}};
\end{axis}   
\end{tikzpicture}
\caption{H versus \nm{\Hp} at \nm{\DeltaT = 0} for various \nm{\Deltap}}
\label{fig:H_Hp__Deltap}
\end{figure}

The dependency of the relationship between the geopotential and pressure altitudes with the temperature offset is graphically represented in figure \ref{fig:H_Hp__DeltaT} for the case without pressure offset (\nm{\Deltap = 0}). In this case, the geopotential and pressure altitudes at both mean sea level and standard mean sea level coincide:
\neweq{\Deltap = 0 \ \longrightarrow \ \HHpzero = \HpHpzero = \HMSL = \HpMSL = 0}{eq:values_Deltap_zero}

Both altitudes diverge as altitude increases in accordance with figure \ref{fig:dHdHp_Hp__DeltaT}. Geopotential altitude H is higher than pressure altitude \nm{\Hp} when \nm{\DeltaT > 0}, while the opposite is true if \nm{\DeltaT < 0}. As the tropopause pressure altitude \nm{\Hptrop} is constant (table \ref{tab:constants}), the corresponding \nm{\Htrop} varies with \nm{\DeltaT}.

Figure \ref{fig:H_Hp__Deltap} graphically represents the dependency with the pressure offset \nm{\Deltap} for the case without temperature offset (\nm{\DeltaT = 0}). As the ratio between increments of both altitudes is unity (figure \ref{fig:dHdHp_Hp__DeltaT}), all lines in the figure are parallel to each other and separated by the difference in their respective \nm{\HpMSL} or \nm{\HHpzero} (horizontal or vertical separation, respectively), which depend on \nm{\Deltap}.
\neweq{\DeltaT = 0 \ \longrightarrow \ H = \Hp - \HpMSL(\Deltap)}{eq:AtmosphereHHP5}
\begin{figure}[h]
\centering
\pgfplotsset{
	every axis legend/.append style={
		at={(0.02,0.75)},
		anchor=west,
	},
}
\begin{tikzpicture}
\begin{axis}[
cycle list={{red,no markers,thick},{orange!50!yellow,no markers,thick},{blue,no markers,thick},{violet,no markers,thick},{green,no markers,thick}},
width=16.0cm,
height=7.0cm,
xmin=0, xmax=15, xtick={0,1,...,15},
xlabel={\nm{\Hp \lrsb{km}}},
xmajorgrids,
ymin=0, ymax=16, ytick={0,2,...,16},
ylabel={\nm{H \lrsb{km}}},
ymajorgrids,
axis lines=left,
axis line style={-stealth},
legend entries={
\nm{\DeltaT = - 20 \lrsb{^{\circ}K}, \, \Deltap = - 5000 \lrsb{kg / m \, sec^{\sss2}}}, 
\nm{\DeltaT = - 20 \lrsb{^{\circ}K}, \, \Deltap = + 5000 \lrsb{kg / m \, sec^{\sss2}}},
\nm{\DeltaT = 0 \lrsb{^{\circ}K}, \, \Deltap = 0 \lrsb{kg / m \, sec^{\sss2}}},
\nm{\DeltaT = + 20 \lrsb{^{\circ}K}, \, \Deltap = - 5000 \lrsb{kg / m \, sec^{\sss2}}},
\nm{\DeltaT = + 20 \lrsb{^{\circ}K}, \, \Deltap = + 5000 \lrsb{kg / m \, sec^{\sss2}}}},
legend columns=1,
legend style={font=\footnotesize},
legend cell align=left,
]
\pgfplotstableread{figs/H_Hp__DeltaT_Deltap.txt}\mytable
\addplot table [header=false, x index=0,y index=1] {\mytable};
\addplot table [header=false, x index=0,y index=2] {\mytable};
\addplot table [header=false, x index=0,y index=3] {\mytable};
\addplot table [header=false, x index=0,y index=4] {\mytable};
\addplot table [header=false, x index=0,y index=5] {\mytable};
\end{axis}   
\end{tikzpicture}
\caption{H versus \nm{\Hp} for various \nm{\DeltaT} and \nm{\Deltap}}
\label{fig:H_Hp__DeltaT_Deltap}
\end{figure}
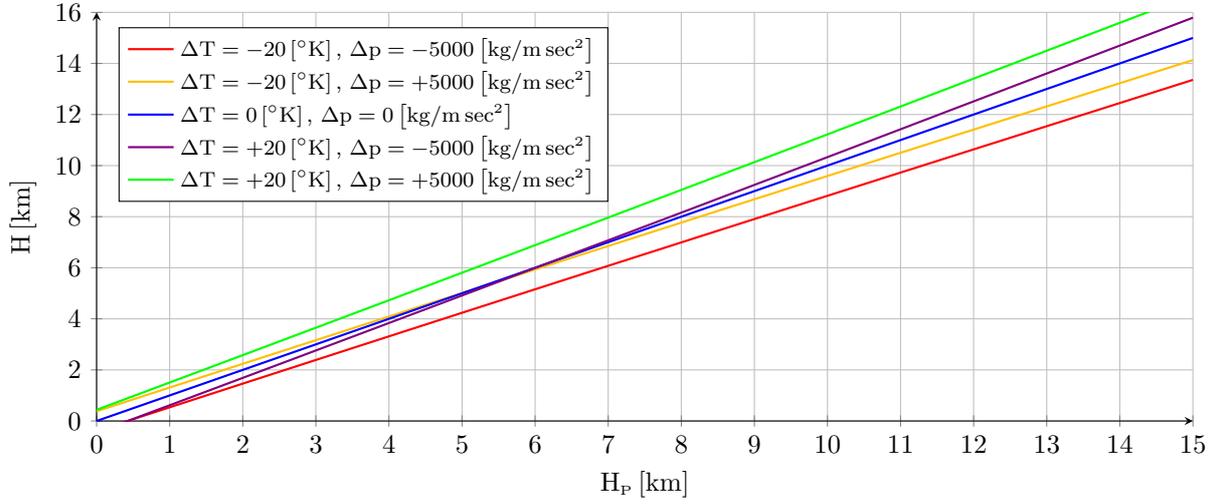

Figure \ref{fig:H_Hp__DeltaT_Deltap} graphically shows the results when neither offset is zero. Atmospheres with the same temperature offset \nm{\DeltaT} are represented with parallel lines, while those with the same pressure offset \nm{\Deltap} intersect the horizontal axis (\nm{H = \HMSL = 0}) at the same point as their \nm{\HpMSL} is the same.

Note that the relationship between geopotential and pressure altitudes can only be explicitly reversed above the tropopause, resulting in (\ref{eq:Hp_H_above}). In the troposphere it is solved by iteration, which converges quickly.
\neweq{\Hpabove = \Hptrop + \frac{\TISAtrop}{\TISA} \; \lrp{\Habove - \Htrop}}{eq:Hp_H_above}


\section{Suggested Use for the Non Standard Atmosphere Model} \label{sec:Suggested}

The ICAO Non Standard Atmosphere or INSA is a quasi static model based on expressions (\ref{eq:model_step1}) and (\ref{eq:model_step2}). The previous sections of this article have focused on the static part of the model (\ref{eq:model_step2}), this is, how to obtain the atmospheric properties at a given geopotential altitude once the temperature and pressure offsets at the current aircraft horizontal position and time have already been determined through (\ref{eq:model_step1}), which is repeated here for clarity.
\neweq{\lrsb{\DeltaT, \, \Deltap}^T = \vec f_1 \lrp{t, \, \lambda, \, \varphi}}{eq:model_step1_bis}

The INSA model does not define the implementation of expression (\ref{eq:model_step1_bis}), instead letting the user customize the functions providing the temperature and pressure offsets so they resemble the conditions encountered by the aircraft in the most accurate way. In this section the author suggests a possible implementation of (\ref{eq:model_step1_bis}) for illustration purposes only, but it is up to the user to implement (\ref{eq:model_step1_bis}) with the required level of realism.

Let's assume without any loss of generality that the trajectory being analyzed corresponds to a flight between two given airports, whose geodetic coordinates are \nm{\vec x_{{\sss GDT},1} = \lrsb{\lambda_1,\,\varphi_1,\,h_1}^T} and \nm{\vec x_{{\sss GDT},2} = \lrsb{\lambda_2,\,\varphi_2,\,h_2}^T}. Let's also consider that the aircraft is expected to depart from \nm{\vec x_{{\sss GDT},1}} at time \nm{t_1} and arrive at \nm{\vec x_{{\sss GDT},2}} at \nm{t_2}.

The temperature and pressure offsets at both airports can be manually set by the user according to the specific conditions to be tested in the simulation. For example, the departing airport \nm{\vec x_{{\sss GDT},1}} and time \nm{t_1} may correspond to a high latitude Northern hemisphere location in winter during the night, which would imply a significantly negative \nm{\DeltaT_1}, while the landing airport \nm{\vec x_{{\sss GDT},2}} and time \nm{t_2} may be those of daytime in the Southern hemisphere close to the tropics, where being summer \nm{\DeltaT_2} should be quite elevated. Similarly, a low pressure weather front may be active at the departure airport, implying a negative \nm{\Deltap_1}, while high pressures may be prevalent at the landing location, resulting in a positive \nm{\Deltap_2}. Note that the altitudes of both airports (\nm{h_1} and \nm{h_2}) do not play any role in the resulting offsets. As explained below, the offsets at both locations may also be determined based on observations taken at both airports at times where the conditions were similar to those expected during the flight.

Once \nm{\DeltaT} and \nm{\Deltap} at both airports have been determined, a simple model for (\ref{eq:model_step1_bis}) would be to have both parameters vary linearly as flight progresses from their values at the departing airport (\nm{\DeltaT_1} and \nm{\Deltap_1}) to those expected when landing (\nm{\DeltaT_2} and \nm{\Deltap_2}). Further realism may be introduced by introducing additional points throughout the flight, so the linear interpolation occurs between the two points closest to the aircraft position. The highest possible accuracy is obtained by employing a grid of temperature and pressure offsets based on longitude, latitude, and time, and using three-dimensional interpolation to get the value of both offsets at each position and time during the flight.


\subsubsection*{Offsets Identification from Ground Observations}

This section shows how to determine the values of the temperature and pressure offsets based on atmospheric pressure and temperature measurements (\nm{\pD, \, \TD}) taken at a given position and time (\nm{\lambdaD, \, \phiD, \, \hD, \, \tD}), usually corresponding to those of an airport or meteorological station, which is assumed to be in the troposphere. A similar approach can be used if the data is taken from a meteorological service such as \cite{NOAA,ECMWF}. The process is the following:
\begin{enumerate}

	\item The location geodetic altitude \nm{\hD} is converted into geopotential altitude \nm{\HD} based on (\ref{eq:h2H}):
	\neweq{\HD = \dfrac{\RE \cdot \hD}{\RE + \hD}}{eq:observe_h2H}

	\item Pressure \nm{\pD} is converted into pressure altitude \nm{\HpD} based on (\ref{eq:Hp_p_below}) and then into standard temperature \nm{\TISAD} based on (\ref{eq:TISA_Hp_below}). Neither conversion depends on the temperature or pressure offsets:
	\begin{eqnarray}
	\nm{\HpD} & = & \nm{\frac{\Tzero}{\betaTbelow} \; \lrsb{\lrp{\frac{\pD}{\pzero}}^{\sss \gBRbelowinv} - 1}}\label{eq:observe_Hp_p_below} \\
	\nm{\TISAD} & = & \nm{\Tzero + \betaTbelow \; \HpD}\label{eq:observe_TISA_Hp_below}
	\end{eqnarray}

	\item The temperature offset \nm{\DeltaTD} is obtained from the difference between the real temperature measurement \nm{\TD} and the standard temperature \nm{\TISAD} based on (\ref{eq:T_Hp_below}):
	\neweq{\DeltaTD = \TD - \TISAD}{eq:observe_T_Hp_below}

	\item The combination of expressions (\ref{eq:TISA_Hp_below}), (\ref{eq:TISAMSL}), and (\ref{eq:H_Hp_below}) results in the following relationship between troposphere standard temperature \nm{\TISAbelow}, mean sea level standard temperature \nm{\TISAMSL}, troposphere geopotential altitude \nm{\Hbelow}, and temperature offset \nm{\DeltaT}. It can be solved to obtain \nm{\TISAMSLD} from \nm{\TISAD}, \nm{\HD}, and \nm{\DeltaTD}.
	\neweq{\Hbelow = \frac{1}{\betaTbelow} \, \lrsb{\TISAbelow - \TISAMSL + \DeltaT \; log_n \, \lrp{\frac{\TISAbelow}{\TISAMSL}}}}{eq:H_TISA}

	\item The mean sea level pressure altitude \nm{\HpMSLD} is obtained from the mean sea level standard temperature \nm{\TISAMSLD} based on (\ref{eq:TISAMSL}):
	\neweq{\HpMSLD = \frac{\TISAMSLD - \Tzero}{\betaTbelow}}{eq:observe_TISAMSL}

	\item The mean sea level pressure altitude \nm{\HpMSLD} is converted into mean sea level pressure \nm{\pMSLD} based on (\ref{eq:p_Hp_below}):
\neweq{\pMSLD = \pzero \ \lrp{1 + \frac{\betaTbelow}{\Tzero} \ \HpMSLD}^{\sss \gBRbelow}}{eq:observe_p_Hp_below}

	\item The pressure offset \nm{\DeltapD} is obtained according to its definition as the difference between the mean sea level pressure \nm{\pMSLD} and the standard pressure at mean sea level \nm{\pzero}:
	\neweq{\DeltapD = \pMSLD - \pzero}{eq:observe_DeltaP}

\end{enumerate}
 
Once both offsets \nm{\DeltaTD} and \nm{\DeltapD} have been identified, the atmospheric properties at any altitude can be computed based on the expressions obtained in this article.


\section{Conclusions} \label{sec:Conclusions}

This article describes an easy-to-use quasi static atmospheric model suited for the requirements of aircraft trajectory prediction and flight simulation. Named the ICAO Non Standard Atmosphere or INSA, it models the possible variations of temperature and pressure with time and horizontal position by means of two parameters, the temperature and pressure offsets. Once these two parameters have been identified for a given location and time, the model provides the variation of the atmospheric properties (temperature, pressure, and density) with altitude complying with all the hypotheses of the ICAO Standard Atmosphere or ISA model \cite{ISA}. The INSA model can be customized so the temperature and pressure offsets adjust to the expected conditions during flight, resulting in more accurate predictions for the atmosphere properties than those provided by ISA. The author has implemented an open-source \nm{\CC} version of the INSA model, available in \cite{Gallo2020_nonstandard_atmosphere}.


\section*{Acknowledgments}

The content of this article is mostly taken from work that the author performed under contract for the European Organization for the Safety of Air Navigation (EUROCONTROL), who possesses all relevant intellectual proprietary rights (\copyright 2021 All rights reserved). The outcome of that work is described in \cite{Gallo2012}, which has restricted distribution. The author would like to thank EUROCONTROL for their permission to publish a section of that work as part of his PhD thesis.

\bibliographystyle{ieeetr}   
\bibliography{insa_atmospheric_model}

\end{document}